# Recent Progress of Heterostructures Based on Two Dimensional Materials and Wide Bandgap Semiconductors

Ying Liu,[1,2] Yanjun Fang,[1] Deren Yang,[1,2] Xiaodong Pi[1,2]*, and Peijian Wang[2]*

[1] State Key Laboratory of Silicon Materials and School of Materials, Science and Engineering, Zhejiang University, Hangzhou, Zhejiang 310007, China

[2] ZJU-Hangzhou Global Scientific and Technological Innovation Center, Zhejiang University, Hangzhou, Zhejiang 311215, China

E-mail: xdpi@zju.edu.cn; pjwang@zju.edu.cn;

**Abstract**

Recent progress in the synthesis and assembly of two-dimensional (2D) materials has laid the foundation for various applications of atomically thin layer films. These 2D materials possess rich and diverse properties such as layer-dependent band gaps, interesting spin degrees of freedom, and variable crystal structures. They exhibit broad application prospects in micro-nano devices. In the meantime, the wide bandgap semiconductors (WBS) with an elevated breakdown voltage, high mobility, and high thermal conductivity have shown important applications in high-frequency microwave devices, high-temperature and high-power electronic devices. Beyond the study on single 2D materials or WBS materials, the multi-functional 2D/WBS heterostructures can promote the carrier transport at the interface, potentially providing novel physical phenomena and applications, and improving the performance of electronic and optoelectronic devices. In this review, we overview the advantages of the heterostructures of 2D materials and WBS materials, and introduce the construction methods of 2D/WBS heterostructures. Then, we present the diversity and recent progress in the applications of 2D/WBS heterostructures, including photodetectors, photocatalysis, sensors, and energy related devices. Finally, we put forward the current challenges of 2D/WBS heterostructures and propose the promising research directions in the future.

# Content

TOC figure:

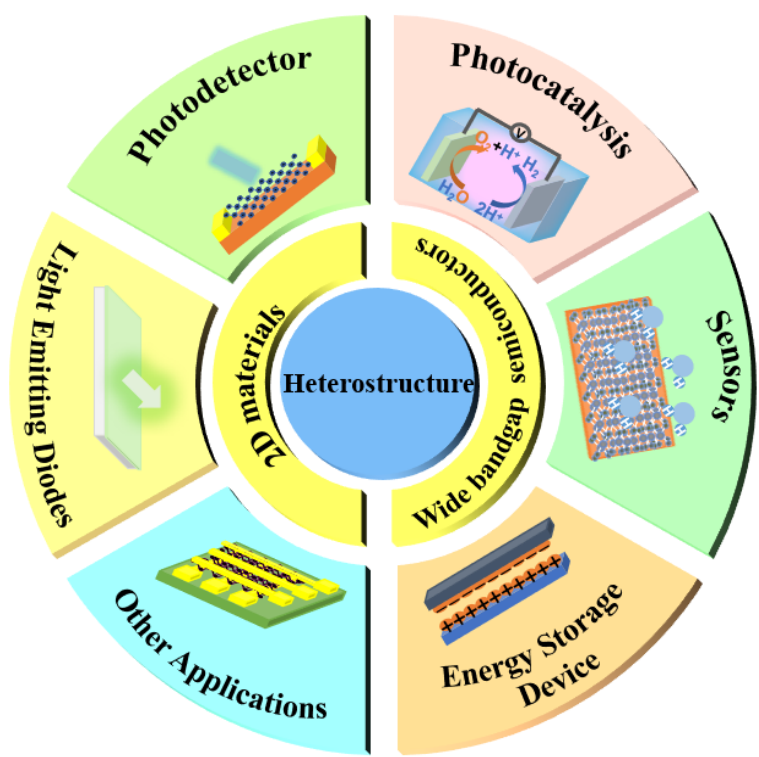

## 1. Introduction

Two-dimensional (2D) materials have been attractive for investigations since the successful isolation of a single atomic layer of graphite material — graphene — by the team of Geim in 2004.[1–3] The 2D materials with atomic-level thickness exhibit large specific surface areas and excellent tunability, significant carrier mobility,[4,5] high electrical conductivity, and excellent optoelectronic properties, etc.[6–8] Also, the van der Waals (vdW) interactions between layers of 2D materials are not restricted by chemical bonding and interfacial lattice matching, so they can be seamlessly integrated on crystalline or amorphous, rigid or flexible substrates.[9–11] In addition to graphene (GN)[1,3], other 2D materials have also attracted widespread attentions, such as the two-dimensional Transition Metal Dichalcogenides (TMDs)[12,13], two-dimensional topological insulators (TI)[14], black phosphorus (BP)[15], hexagonal boron nitride (hBN)[16], MXene[17], etc. The superior properties and rapid development of the 2D materials family has greatly boosted its performance and applications in electronic and optoelectrical devices[18,19], including photodetectors from ultraviolet to infrared[20–22], field effect transistors[23–25], memristor[26,27], and novel energy devices[28,29]. Although 2D materials show excellent performance and application potential, there is still a need to adjust their electronic and optoelectronic properties to accommodate the applications in various scenarios. Therefore, the idea of assembling different 2D materials into heterostructures to improve their properties was born, and heterostructures based on 2D materials are expected to yield unprecedented properties.[30–32]

Wide bandgap semiconductors (WBS), such as silicon carbide (SiC), gallium nitride (GaN), and diamond, display large energy bandgap more than 3 eV, strong breakdown electric field, and high electron mobility, as shown in Table 1.[33–38] In the past few years, significant progress has been made on the two typical WBS: SiC[39] and GaN[34], which exhibit great potential for applications in high-temperature, high-pressure, high-power and high-frequency electronic devices due to their excellent thermal and electronic properties.[40–43] For SiC and GaN, the potential of application stems from its excellent physical and electronic properties. As shown in Table 1, the large band gap enables the device to operate at high temperature and low leakage currents, the high breakdown electric field and large electron saturation velocity enable SiC to generate high power at high frequencies. Among the different polytypes in the crystal structure of SiC, the 4H-SiC receives more attention than other polytypes due to its large band gap of 3.26 eV and electron mobility around 900 $cm^2 \cdot V^{-1} \cdot s^{-1}$ (Table 1).[39] Besides, the development of optically active spin quantum bits in SiC provides a solution for the application of quantum repeaters and the development of scalable quantum networks.[44] According to the material properties of GaN in Table 1, the direct wide band gap (3.4 eV) with strong atomic bonds, thermal conductivity of 2.1 W/cm, strong resistance to radiation and good chemical stability,[43] enabling a wide range device applications of GaN in the market, including power electronics and deep ultraviolet photodetectors.[42,43] As the promising semiconductor materials, SiC and GaN are the foundation of the emerging semiconductor industry, and widely used in electronic and optoelectronic devices.[33,45,46]

Table 1. Material properties of SiC, GaN, and diamond.[33–36]

| Property | 4H-SiC | GaN | Diamond |
|---|---|---|---|
| Wide band gap | 3.26 eV | 3.4 eV | 5.49 eV |
| Breakdown electric field | 5 MV/cm | 3.5 MV/cm | 200 MV/cm |
| Dielectric constant | 9.7-10.0 | 9.5 | 5.5 |
| Electron mobility | 900 cm$^2$/V·s | 1300 cm$^2$/V·s | 2400 cm$^2$/V·s |
| Electron saturation velocity | $2\times10^7$ cm/s | $2.5\times10^7$ cm/s | $1.5\text{-}2.7\times10^7$ cm/s |
| Thermal conductivity | 3.5 W/cm | 1.3–2.1 W/cm | 20 W/cm |

In 2D materials, the variety of bandgap is a zoo of metal (NbS$_2$, TaS$_2$ etc.),[47] semimetal (graphene, WTe$_2$ etc), narrow bandgap (TMDs such as MoS$_2$, WS$_2$, MoSe$_2$, WSe$_2$), as shown in Figure 1. Combining with WBS, the bandgap can be extended to the ultraviolet region or even further in the material system (Figure 1).[48–50] Furthermore, the superior lattice and thermal expansion registry etc. (will be discussed in the following) determine this system to be an interesting ensemble. With the increasing interests in the research of 2D materials, there have been many reviews and reports on 2D heterostructures in recent years,[5,31,32,51,52] but currently there is no systematic description of SiC- or GaN-based 2D heterostructures. The fascinating properties of these heterostructures, such as the dangling-bond-free surface and the tunable charge depletion layer and band gap etc., igniting research interests in new physical explorations. Besides, the good lattice match of 2D/WBS heterostructures is expected to improve the crystal quality for practical devices. It is necessary and timely to provide an overview of recent developments in 2D/WBS heterostructures and their devices, considering the rapid progress and promise in the research area.[5] Here, we focus on the heterostructure of 2D materials and the two most commonly used WBS: SiC and GaN, which has not been discussed in detail by other reviews yet currently. This will be beneficial to advance the research and development in this realm. The main fabrication methods and progress of 2D/WBS heterostructure in different application fields are introduced. We summarized the potential applications of 2D/WBS heterostructures in photodetectors, photocatalyst, sensors, light emitting diodes (LEDs) and energy storage devices, etc. Then we summarized the article and put forward the main obstacles to overcome in the future. This work provides a very useful reference for the future direction for development of the heterostructure of 2D materials based on SiC and GaN.

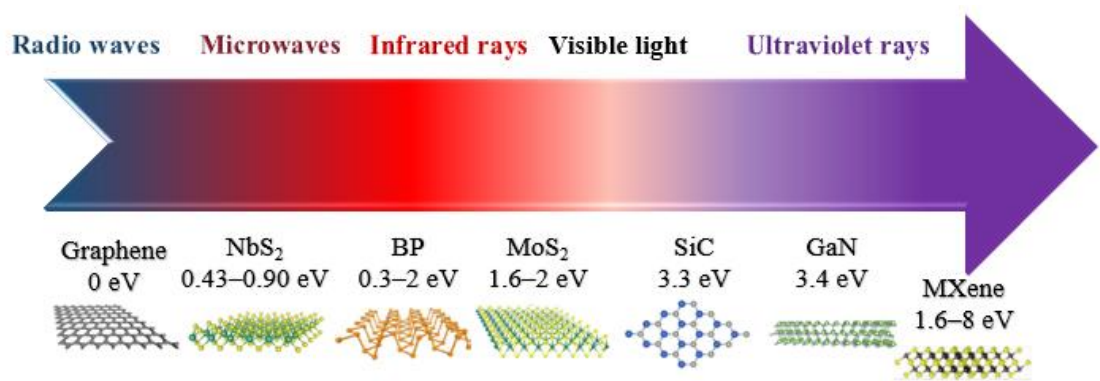

Figure 1 The spectrum of the bandgap of graphene, $NbS_2$, BP, $MoS_2$, MXene, SiC and GaN. Reproduced with permission.[48][49][53][54][55]

## 2. Preparation of 2D/WBS Heterostructures

For 2D materials grown on WBS, different layers are combined through weak vdW interactions for perfect 2D materials, and therefore constraints such as chemical bonding and lattice matching are relaxed. However, epitaxy of 2D materials on WBS is governed by similar considerations, such as lattice constant matching for conventional heteroepitaxial systems. In reality, chemical bonding often exists, because of defects and vacancies in the interfaces. The vdW layer also orients to minimize energy concerning electrostatic interactions and orbital overlap. Similarly, the strain and morphology of a vdW 2D material on the substrate are arising from the competition between the energy of film-substrate vdW interactions and the elastic energy associated with distortion of the layer.[56] Table 2 shows the properties of lattice type, symmetry, lattice constants and thermal expansion coefficients of typical 2D materials and WBS. For example, $MoS_2$ possesses the same lattice symmetry with GaN, and the lattice constant mismatch is very small (< 1%), compared with conventional substrate like silicon (~40%). So are the good lattice matching of $MoS_2$/SiC and $MoSe_2$/GaN, which causes the monolayer domains to align with the rotation symmetry, reducing the grain boundaries of heterostructure.[57] Compare the symmetry and adaptation of 2D materials with WBS and silicon (Si), the diamond structure Si exhibits higher lattice mismatch and greater discrepancy on thermal expansion coefficient with $MoS_2$ or $MoSe_2$, while hexagonal polytype GaN or SiC show better lattice match and small thermal expansion coefficient mismatch with 2D materials for the stability of epitaxial alignment. The matching of symmetry, lattice constant and the thermal expansion coefficients endows the WBS a promising candidate for an ideal interface in the epitaxial growth of the 2D materials.

Table 2 Symmetry, in-plane lattice constant *a*, and thermal expansion coefficients of 2D materials and 3D WBS materials at room temperature.

| Material | Structure | Space group | Lattice constant a | Thermal expansion coefficient in *a* | Ref |
|---|---|---|---|---|---|
| Graphene | hexagonal | *P6/mmm* | 2.46 Å | $-7 \times 10^{-6}$ K$^{-1}$ | [58][59][60] |
| MoS$_2$ | hexagonal | *P3m1* | 3.16 Å | $4.92 \times 10^{-6}$ K$^{-1}$ | [61][57] |
| MoSe$_2$ | hexagonal | *P3m1* | 3.28 Å | $7.24 \times 10^{-6}$ K$^{-1}$ | [61][62] |
| 4H-SiC | hexagonal | *P6$_3$/mc* | ~3.10 Å | $3.21 \times 10^{-6}$ K$^{-1}$ | [63][64][65] |
| GaN | hexagonal | *P6$_3$/mc* | 3.19 Å | $3.95 \times 10^{-6}$ K$^{-1}$ | [57][63] |
| Si | cubic | *Fd3m* | 5.43 Å | $2.45 \times 10^{-6}$ K$^{-1}$ | [66][67] |

The fabrication and reliable integration techniques of 2D/WBS heterostructures are of great importance for further research and device applications. Currently, the methods to realize 2D/WBS heterostructures are divided into three approaches, including exfoliation and transfer,[68,69] direct preparation methods such as Chemical Vapor Deposition (CVD)[70–72] and Pulsed Laser Deposition (PLD)[73–75], and the combination methods of physical or chemical approaches with exfoliation and transfer.[76]

By optimizing the exfoliation and technologically scalable transfer technology, high-quality 2D heterostructures can be effectively obtained. The exfoliation including mechanical exfoliation[77], liquid phase exfoliation[2], and electrochemical exfoliation[78], have frequently been used to explore emerging 2D materials. The most commonly used approach is the tape peeling technique from bulk materials. Gao et al. exfoliated the large-area and high-quality 2D monolayers by a gold adhesion layer with covalent-like quasi bonding to the layered crystal, as shown in Figure 2a–d, the large-area monolayer MoS$_2$ was exfoliated onto the Au (2 nm)/Ti (2 nm)/SiO$_2$/Si substrate by tape.[79] A variety of monolayer 2D materials can be obtained with this Au-assisted exfoliation, and the universal method can be extended to the exfoliation of 40 types 2D monolayers and transfer to rigid or flexible substrates. Similarly, the method can be used for the fabrication of 2D heterostructures, such as the MoS$_2$/WSe$_2$ heterostructure shown in Figure 2e–k. Large-area MoS$_2$ and WSe$_2$ monolayers were exfoliated onto two separate Au/Ti/SiO$_2$/Si substrates, respectively. The Au film was then etched with KI/I$_2$ solution, and the monolayer MoS$_2$ was transferred onto WSe$_2$ with PMMA, forming the MoS$_2$/WSe$_2$ heterostructure. Moreover, Li et al. obtained the MoS$_2$ samples *via* mechanically exfoliated from bulk MoS$_2$ crystals onto Si/SiO$_2$ substrates with alignment marks, then transferred the multilayer MoS$_2$ to doped p-GaN substrate with 4 nm Al$_2$O$_3$.[80] The vertically stacked MoS$_2$ heterostructure exhibited the broad-area electroluminescence (EL) emission throughout the junction area and a strong EL emission in multilayer MoS$_2$ with indirect bandgap.

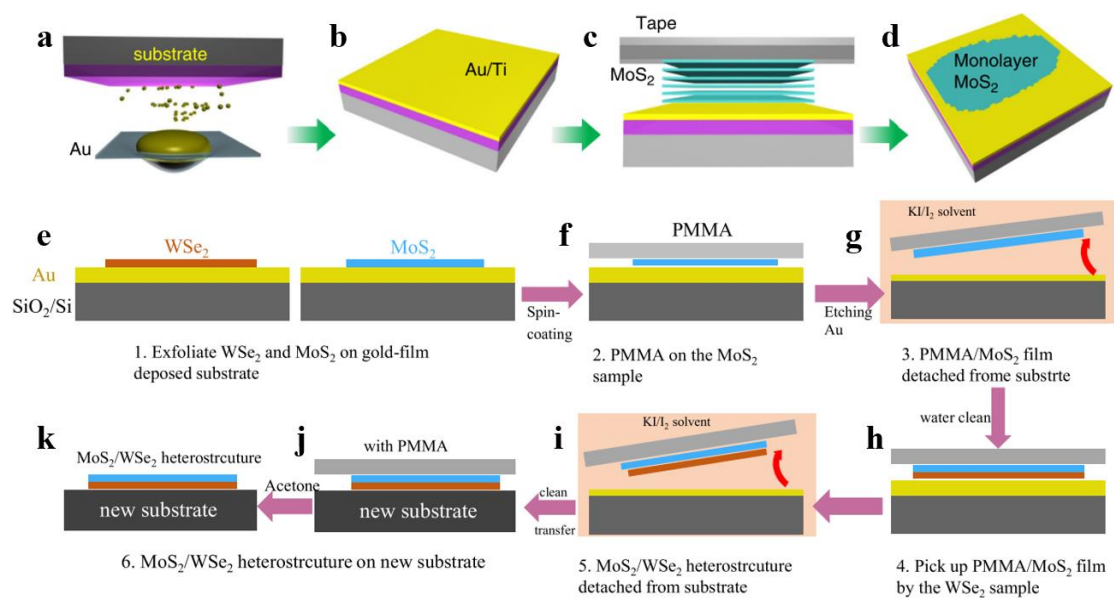

Figure 2 Schematic of exfoliation process of monolayer MoS$_2$. a. The Au film were deposited by electron evaporation on Si substrate with an adhesion metal layer (Ti or Cr), b. the Au/Ti was obtained on substate, c. layered crystal MoS$_2$ was cleaved from tape and placed on Si, d. the tape was pressed vertically and removed to obtain a large-area monolayer MoS$_2$ flake. Fabrication process of the MoS$_2$/WSe$_2$ heterostructure by exfoliation and transfer. e. The WSe$_2$ and MoS$_2$ monolayers were exfoliated on two separate Au/Ti/SiO$_2$/Si substrates, f. PMMA was spin-coated on MoS$_2$, g. the sample was put into KI/I$_2$ solution to etch the Au film and obtain the PMMA/MoS$_2$ flake, h. the PMMA/MoS$_2$ was washed in DI water and picked up using WSe$_2$ sample, then baked at ~100 ℃ to ensure the contact between two layers, i. the WSe$_2$/MoS$_2$ sample was put into KI/I$_2$ solution to etch the Au film, j. the hetero-bilayer sample with PMMA was picked up using new substrate, k. PMMA was removed with acetone and the WSe$_2$/MoS$_2$ heterostructure was obtained. Reproduced with permission.[79]

CVD is a widely used method to prepare 2D materials, which can obtain wafer scale thin films.[81] Ding et al. grew graphene films directly on GaN/sapphire substrates by CVD under the catalyst-free, atmospheric pressure and CH$_4$/Ar ratio of 5:100.[59] The obtained films were mostly bilayer graphene, with a small portion of monolayer. Chu et al. successfully synthesized MoSe$_2$ films on GaN with the traditional CVD process, and grew the several layer MoSe$_2$ films with controlled lattice orientation by selenizing the MoO$_3$ powder.[82] The combination of CVD, exfoliation and transfer method is often used to obtain the 2D/WBS heterostructures, as shown in Figure 3.[83,84] Chen et al. synthesized vertical heterostructures of p-MoS$_2$/n-MoS$_2$ with a two-step CVD process, and transferred the p-MoS$_2$/n-MoS$_2$ to as-grown p-GaN/sapphire substrate.[85] Additionally, Liao et al. grew doped GaN films on sapphire substrate with metal-organic chemical vapor deposition (MOCVD), then synthesized the monolayer graphene by CVD, and transferred the graphene to GaN with poly(methyl methacrylate) (PMMA) to obtain graphene/GaN heterostructure.[86] Similarly, Journot et al. grew graphene on copper foil by CVD, and transferred it to a sapphire substrate with wet transfer. They then grew GaN microstructures on graphene in a commercial MOCVD system with a closed coupled nozzle system to obtain the heterostructure.[87]

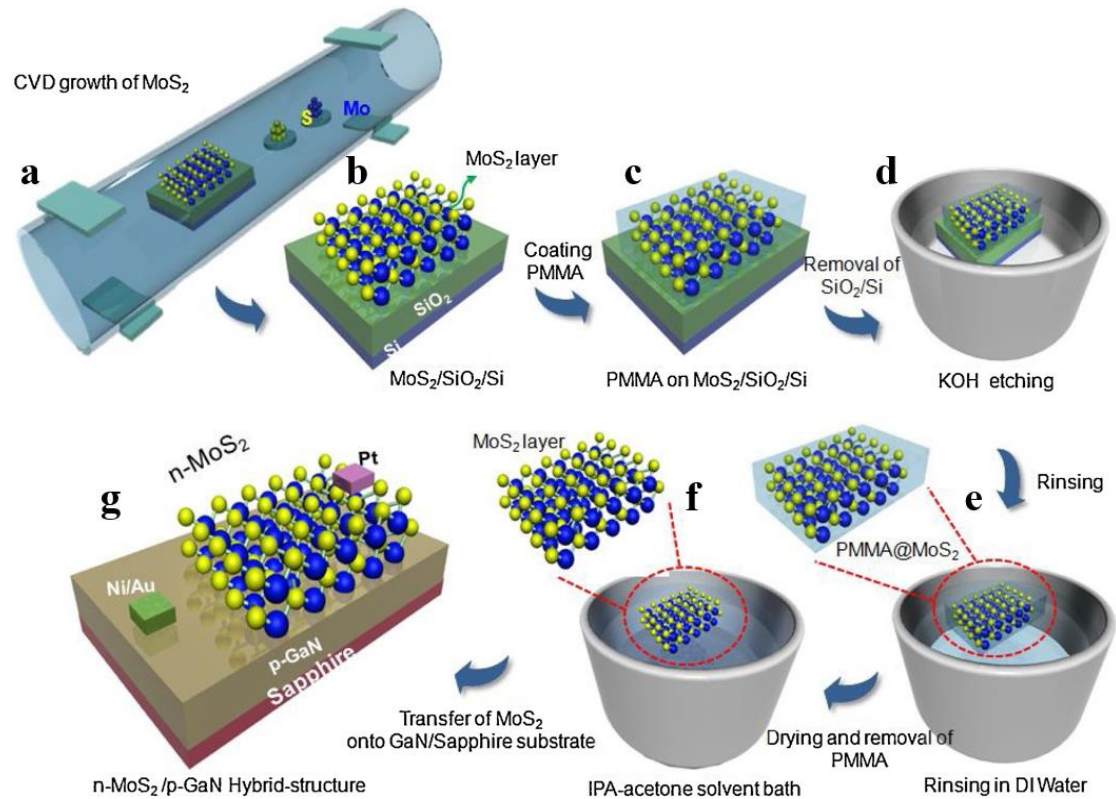

Figure 3 The preparation of n-MoS$_2$/p-GaN heterostructure devices. a. CVD setup; b. n-MoS$_2$ layer was grown on SiO$_2$/Si; c. Spin coating PMMA on MoS$_2$/SiO$_2$/Si; d. Wet etching the substrate in KOH solution; e. Rinse the separated sample in deionized water and dry for 24 hours; f. Detach PMMA in isopropyl alcohol (IPA)-acetone solution; g. Mechanical transfer of MoS$_2$ onto p-GaN/sapphire. Reproduced with permission.[84]

Physical vapor deposition (PVD) method for growing 2D heterostructures, especially PLD, has been an easy but not commonly explored method. Serrao et al. prepared large-area, highly crystalline and out-of-plane textured MoS$_2$ films on Al$_2$O$_3$,

GaN and 6H-SiC substrates by PLD.[88] Yu et al. used PLD technique to deposit 2 nm thick layered 2D MoS$_2$ on sapphire substrate, then used plasma-assisted molecular beam epitaxy (PAMBE) system to grow GaN films, and prepared GaN/MoS$_2$ heterostructures at different temperatures.[89] Also, the combination of physical and chemical methods with exfoliation and transfer is used to prepare the 2D/WBS heterostructures. Kumar et al. grew a 300 nm GaN film on sapphire substrate with molecular beam epitaxy (MBE), then deposited the Mo film on GaN by magnetron sputtering, and finally sulfurized the Mo film to grow MoS$_2$ by CVD.[90,91] They also used the metal-organic vapor phase epitaxy (MOVPE) method to grow epitaxial GaN films on sapphire substrates, then transferred MoS$_2$ flakes from MoS$_2$ crystals to GaN/sapphire with the mechanical exfoliation method and obtained the MoS$_2$/GaN heterostructures.[92] Besides, Herman et al. prepared GaN films on Mg-doped p-type and Si-doped n-type GaN/sapphire by MOVPE and PAMBE, then grew monolayer graphene on copper foil *via* PECVD, and transferred the graphene with PMMA to GaN/sapphire semiconductors.[93]

Numerous research efforts have reported on the combination of physical and chemical approaches or etching methods to integrate 2D/WBS heterostructure devices. For example, Zhao et al. prepared the monolayer graphene/nanoporous GaN heterostructures for highly sensitive ultraviolet (UV) photodetectors.[94] The nanoporous GaN has a high surface-to-volume ratio, which can increase the density of surface state and reduce the Schottky barrier height of the interface by trapping the photocarriers (holes) at the interface, resulting in higher sensitivity. First, they grew the GaN epitaxial layers by MOCVD, and converted the bulk GaN epitaxial layers into nanoporous GaN with photo-assisted electrochemical etching. Next, the high-quality monolayer graphene films were grown on copper foil by CVD, and transferred to SiO$_2$/Si substrate. After that they spin-coated the substrate with PMMA, and patterned the PMMA/graphene/SiO$_2$/Si by electron beam lithography. Then, they etched the substrate with oxygen plasma, and consequently obtained the PMMA/graphene films. Finally, the PMMA/graphene microsheets were transferred onto nanoporous GaN sheets with a site-specific transfer printing technique, and the PMMA was removed by acetone to obtain the graphene/nanoporous GaN heterostructures.

In addition, the construction of 2D/WBS heterostructure materials can be achieved with many other methods, such as solution processing method, wet chemical method or their combinations.[51,95] Devi et al. designed hexagonal trap structures on p-doped GaN by wet chemical etching with molten KOH, and then grew MoS$_2$ nanowall networks directly on p-GaN *via* a thermal solvent technique.[96] Liu et al. synthesized 2D-GaN flakes by liquid metal printing and surface-constrained nitridation reaction method. They then used transparent tape to peel p-MoS$_2$ flakes from commercial MoS$_2$ material onto polydimethylsiloxane (PDMS) films, and transferred to the prepared n-GaN flakes for MoS$_2$/GaN vdW heterostructure.[97] Besides, Gao et al. prepared the Ti$_3$C$_2$T$_x$/GaN vdW heterostructures with a combination method, where T$_x$ represents the functional groups such as -O, -OH and/or -F groups, and the fabrication process is shown in Figure 4.[98,99] First, they selectively etched the MAX phase precursors (Ti$_3$AlC$_2$ powder) in HCl/LiF aqueous mixture, and got the 2D Ti$_3$C$_2$T$_x$ nanosheets. Next, they dissolved the Ti$_3$C$_2$T$_x$ film in deionized water, and obtained Ti$_3$C$_2$T$_x$ colloidal solutions with various concentrations. Then, they cut the polyethylene glycol terephthalate (PET) tape on the GaN substrate into square holes with a laser engraving machine, and drop-cast the Ti$_3$C$_2$T$_x$ colloidal solutions on the n-type or p-type GaN substrates. Finally, they naturally dried the Ti$_3$C$_2$T$_x$ colloidal solutions at room temperature, and formed the Ti$_3$C$_2$T$_x$/(n/p)-GaN van der Waals heterostructures.

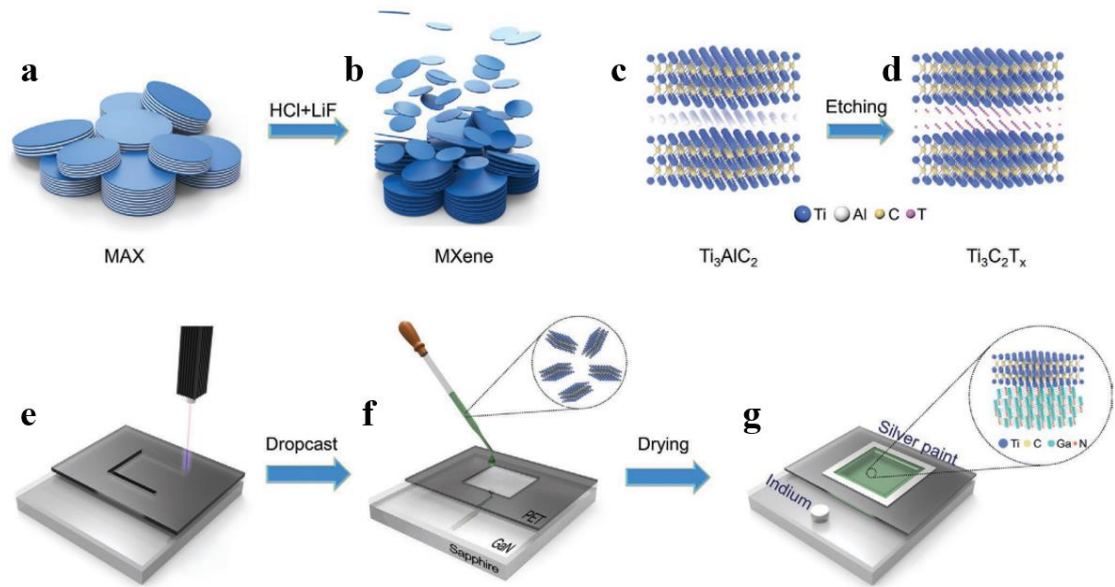

Figure 4 Schematic diagram of the fabrication of Ti$_3$C$_2$T$_x$/GaN heterostructure devices. a. MAX phase material; b. MXene was obtained from MAX by aqueous mixture of HCl/LiF; Crystal structure of c. Ti$_3$AlC$_2$ and d. Ti$_3$C$_2$T$_x$ nanosheets; e. 3 × 3 mm$^2$ square hole engraved on glycol terephthalate (PET) tape; f. Ti$_3$C$_2$T$_x$ colloidal solution was drop-cast in the hole and dried at room temperature; g. The edges of the square hole were coated with silver glue and Indium was used as the electrode. Reproduced with permission.[99]

## 3. Application of 2D Heterostructures Based on SiC

The WBS material SiC,[39] exhibits an excellent combination of physical and electronic properties, such as high thermal conductivity, high breakdown electric field, and large electron saturation drift rate, showing great potential for applications in high frequency and high-power devices.[41] The good chemical stability of SiC-based 2D heterostructures is utilized to make devices with outstanding stability in harsh environments. The large band gap of SiC enables the devices to operate at high temperature with low leakage current. Fang et al. investigated the effect of hydrogenated SiC sublayers (unit cell in vertical orientation **c**) at the interface of heterostructure on electronic states and charge transfer in monolayer $MoS_2$ by first-principle calculations.[64] It was found that the monolayer $MoS_2$ exhibited metallic properties or ambipolar-type polarization when in contact with the hydrogenated Si- or C- terminations of H-SiC ($MoS_2$/H-SiC or $MoS_2$/SiC-H). According to the calculation, the electronic state of monolayer $MoS_2$ on SiC sheet (modeled in the thickness of unit cell) was significantly tuned by hydrogenation, potentially generating the p-n junctions in 2D heterostructures, which is essential for electronic device applications.[64,100]

Heterostructure of 2D material with SiC which possesses the above-mentioned superior properties could generate new physical behaviors and interesting functionalities. In the following, several kinds of electronic and optoelectronic applications will be reviewed for the integration of 2D materials with SiC.

**3.1 Photodetectors**

Photodetectors have promising applications in a variety of fields including image sensing, communications, environmental monitoring, and space exploration. SiC is considered to be one of the most important WBS materials in the development of UV photodetectors.[101] The 2D/SiC heterostructure photodetectors could effectively improve factors of merits such as sensitivity, spectral selectivity, responsivity and detectivity, external quantum efficiency (EQE) of photodetectors, due to the modulation of charge carrier-dynamics at the interface of heterostructure.[102][103]

The photodetectors of graphene/SiC heterostructures were mostly investigated. For example, Yang et al. demonstrated a heterostructure Schottky UV photodiode with epitaxial graphene (EG) grown on n-type C-face 6H-SiC that can be used for UV photodetection.[104] The results showed the response speed of the device in the range from tens of nanoseconds to milliseconds. The responsivity reached 2.18 A/W at 5 V with a 325 nm laser, and the EQE increased to more than 800% due to the trap-induced internal gain mechanism. The traps were verified by studying the time constant as a function of UV laser power, and the monotonically decreasing time constant with increasing laser power at certain biases suggested the existence of traps in SiC substrate. Under the high reverse bias voltage, photocarriers rapidly separated to anticipate photoconduction, and the photoelectrons in the conduction band might be captured by the electronic traps, so that the photogenerated carriers recycle multiple times in the circuit before recombination, achieving an internal gain mechanism related to the trap-assisted photocurrent multiplication. Besides, Li et al. reported the UV photodetector with high responsivity and low operating voltages based on graphene/4H-SiC (Si face) wafers, with the structure schematic shown in Figure 5a.[105] They realized the electric doping of graphene layer by varying the gate voltage, and injecting photogenerated carriers from SiC with 325 nm laser excitation to achieve the photodoping of graphene layer. Under negative gate-source voltage $V_{GS}$, the photogenerated holes were injected into the channel and the channel exhibited p-type, resulting in a planar graphene n-p-n junction. Under positive $V_{GS}$, the channel exhibited an overall n-type with strong electron doping, resulting in a planar graphene n-n-n junction. Note n-p-n and n-n-n are just the notation to represent the carrier type in different area of the graphene. As shown in Figure 5b, the photoresponsivity reached 254.1 A/W at source-drain voltage $V_{DS}$ = −3 V, $V_{GS}$ = 3 V due to the high photoconductivity gain of the planar n-n-n junction, and the calculated maximum value of EQE is $9.6 \times 10^4$% at $V_{DS}$ = −3 V. Such high performance is attributed to the high photogenerated carrier concentration due to light doping, assuming that all the obtained photons are absorbed by 4H-SiC and excite photogenerated carriers by 100%. In addition, the greatly extended electron lifetime contributes significantly to the high photoconductivity gain due to the complex carrier exchange between graphene and SiC, resulting in the high responsivity and EQE. In Figure 5c and 5d, the low photo-response time and high photocurrent under different $V_{DS}$ and $V_{GS}$ demonstrated the improvement of photoconductivity. Due to the dual modulation of optical and electric fields, the negative photoconductivity improved under negative gate voltage and normal photoconductivity facilitated under positive gate voltage. The metal-graphene-metal UV photodetector was achieved by fabrication of EG/SiC heterostructure on semi-insulating 4H-SiC, providing a reference for the design of other graphene/SiC heterostructure optoelectronic devices.[106] The EG/4H-SiC (Si face) heterostructure photodetector exhibited photogenerated electron transfer and then carrier recombination under illumination. The photogenerated electrons in SiC migrate to graphene under built-in electric field, where the carrier recombination occurs. Single layer graphene channel (SL-PD) with an aspect ratio of 100 exhibited an on/off ratio of 727.8%, and an accelerated response rate with rise time $t_r$ ~ 5 ms and decay time $t_d$ ~ 15 ms. In addition, the SL-PD displayed a stable photo-response signal and EQE of 189.7 mA/W and 66.6 %, respectively.

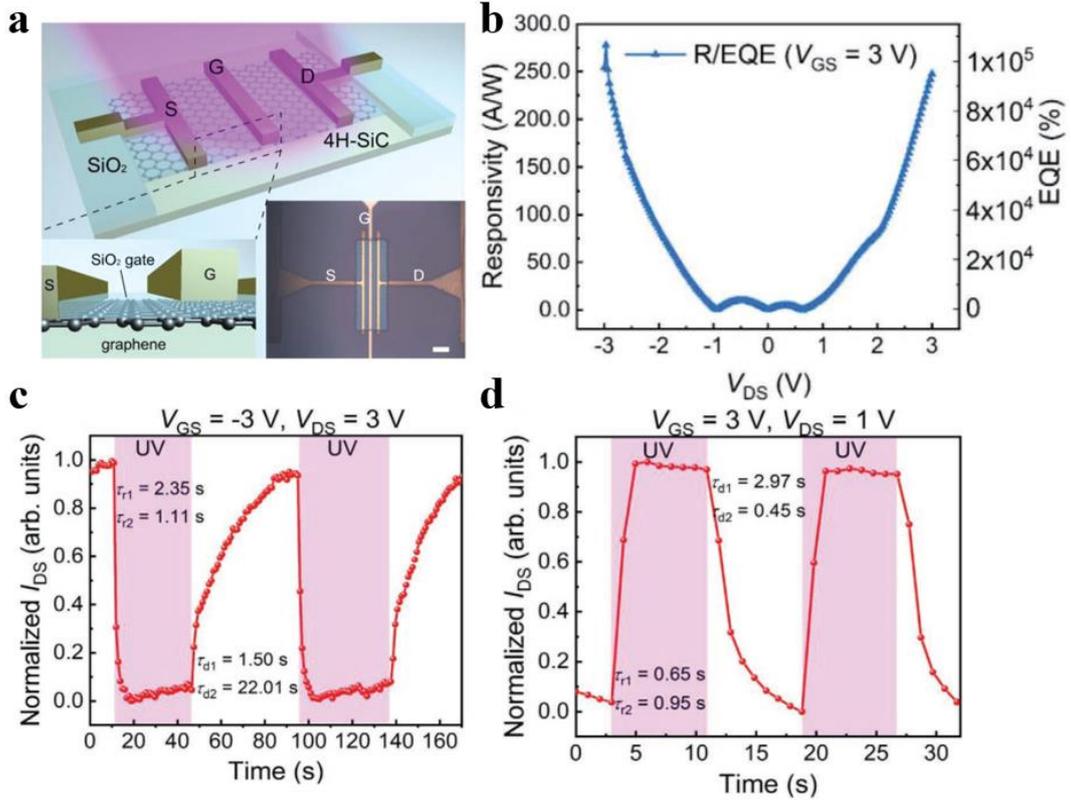

Figure 5 a. Structure schematic diagram and optical microscope image of the device. b. Calculated responsivity (R) and EQE at $V_{GS}$ of 3 V under different gate voltages. Normalized transient photocurrent $I_{DS}$ under different time condition with c $V_{GS}$ = −3 V, $V_{DS}$ = 3 V and d $V_{GS}$ = 3 V, $V_{DS}$ = 1 V. Reproduced with permission.[105]

The range of photodetector response of 2D/SiC heterostructure materials can reach from UV region to the visible region. Xiao et al. demonstrated a $MoS_2$/4H-SiC (Si face) photodetector synthesized by direct CVD with excellent performance in UV and visible regions.[107] The detector showed a low noise equivalent power of $10^{-13} \sim 10^{-15}$ W/Hz$^{1/2}$, typical response time of rise time $t_r$ (defined as the time required for photodetector to increase from 10% to 90% of the final output) ~ 0.58 s and decay time $t_d$ (the time required for photodetector to decrease from 90% to 10% of the trailing edge of the pulse waveform) ~ 0.33 s, and responsivity up to 5.7 A/W under UV irradiation. Moreover, Gao et al. investigated the TMDs $MoSe_2$/$WSe_2$ heterostructure photodetector based on epitaxial n-doped 4H-SiC which provided enhanced photo-response behavior through gate modulation, exhibiting low leakage current, high stability and fast photo-response.[108] The maximum responsivity of the heterostructure photodetector was 7.17 A/W, the corresponding maximum EQE and detectivity were $1.67 \times 10^3$% and $5.51 \times 10^{11}$ Jones with a maximum $I_{light}/I_{dark}$ ratio of ~$10^3$. Many electrons in this n-type heterostructure can participate in the photocurrent, increasing the carrier concentration at the gate voltage, and resulting an EQE much high than 100%.

**3.2 Photocatalysis**

Photocatalysis, which can directly convert solar energy into chemical energy, is recognized as one of the most effective methods to solve the problems of energy shortage and environmental pollutions.[109] Currently, the low separation efficiency and mobility of photogenerated carriers are the main limitations of photocatalysis. Heterostructures of 2D materials and SiC, with type II energy band arrangement such as $MoS_2$/SiC and boron phosphide/SiC, can separate electrons and holes at the interface more efficiently, making them excellent candidates for photocatalysts.[110,111]

Mathur et al. proposed a new method to modulate photocatalytic activity and band gap narrowing in EG/SiC (6H-SiC and 3C-SiC) to produce efficient photocatalysts.[112] The photocatalytic reaction rate for EG/6H-SiC composites were improved by ~1000% under UV light, and the bandgap was narrowed to 2 eV or even smaller under the visible light with a tunable quality factor $I_{2D}/I_G$. The Raman intensity ratio $I_{2D}/I_G$ of the 2D band to the G band of graphene played a crucial role in modulating the band gap and enhancing the photocatalytic activity of EG/SiC composites. The suitable selection of $I_{2D}/I_G$ could provide excellent photocatalytic activity under UV light. The photocatalytic degradation mechanism was shown in Figure 6a, the photogenerated electrons (e⁻) were excited from value band (VB) of 6H-SiC to conduction band (CB), then transferred to the Fermi level of EG, and reacted with $O_2$ molecules to form reactive oxygen species (•$O^{2-}$) which can oxidize Rhodamine B molecules. The enhanced photocatalytic performance of this EG/SiC composite was attributed to the efficient transfer of photogenerated electrons at the interface, and the prompt separation of photogenerated electrons or holes of high-quality graphene. Furthermore, the large difference between the conduction band of SiC and the Fermi energy level of graphene allowed the rapid transfer of photogenerated electrons from SiC to graphene, avoiding the recombination of electrons and holes in SiC, which significantly enhanced the efficiency of charge separation.[113]

In addition, Wang et al. introduced $Ti_3C_2$ MXene quantum dots into SiC to construct a novel heterostructure catalyst.[115] The composite exhibited excellent photocatalytic $NO_x$ removal performance under visible light irradiation. The NO pollutant removal efficiency was 74.6%, which was 3.1 and 3.7 times higher than that of pure $Ti_3C_2$ MXene quantum dots and SiC, respectively. Liao et al synthesized another photocatalyst based on heterostructure of $SnO_2$/3C-SiC nanowires.[114] The schematic diagram of photocatalytic hydrogen precipitation mechanism is shown in Figure 6b. In detail, under simulated sunlight, electrons were excited from VB to CB, and the same number of holes were generated in VB. Since $SnO_2$ has a more positive conductive band edge than SiC, the photogenerated electrons are injected from CB of SiC into CB of $SnO_2$, and holes are injected in the opposite direction to achieve accelerated separation of electron-hole pairs and inhibit recombination. The highest hydrogen evolution rate of 274 μmol·g$^{-1}$·h$^{-1}$ under mixed light was about 4 times higher than that of pristine SiC nanowires, and the current density of 62.0 mA/cm$^2$ at 0.6 V of SnO$_2$/SiC nanowire cathode was 6.9 times higher than that of pristine SiC nanowires photoelectrode.

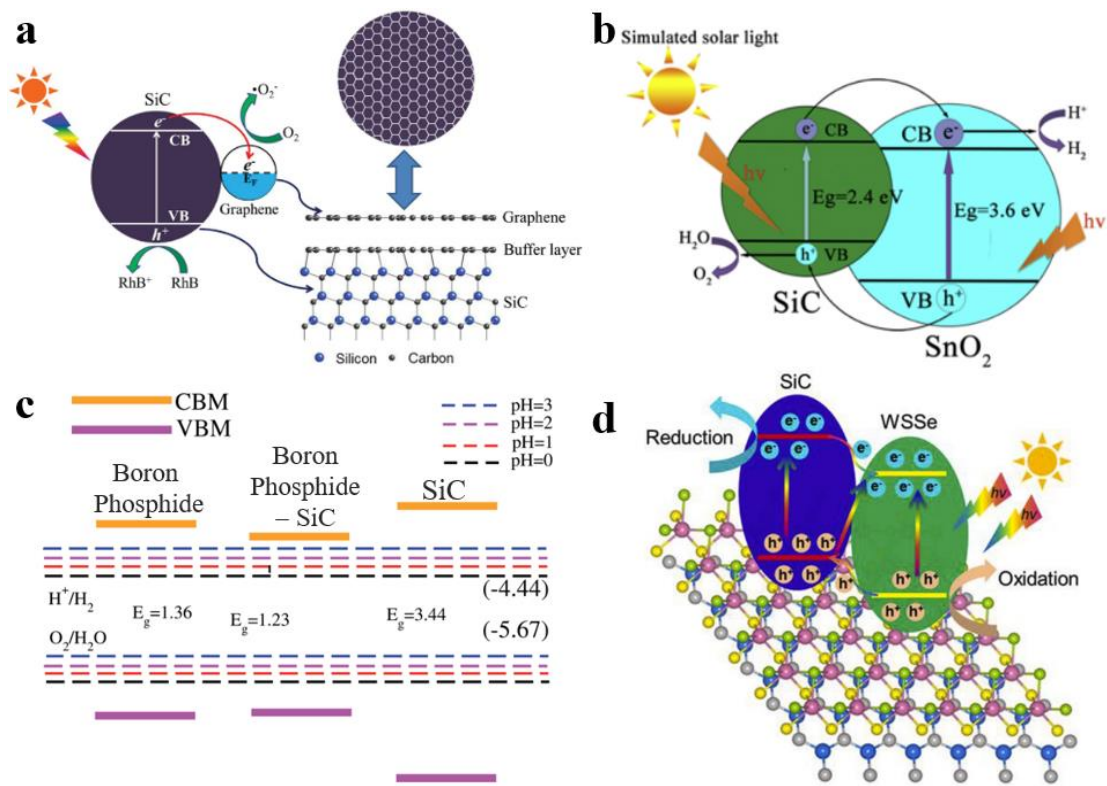

Figure 6. a. Scheme of the photocatalytic degradation mechanism of Rhodamine B in EG/6H-SiC composite under UV illumination. Reproduced with permission.[112] b. Schematic of charge separation in SnO$_2$/3C-SiC NWs. Reproduced with permission.[114] c. Photocatalytic response of boron phosphide, SiC, and their vdW heterostructure. Reproduced with permission.[111] d. Schematic of photogenerated electrons and holes migration at the WSSe/SiC interface. Reproduced with permission.[55]

Moreover, Do et al. designed the vdW heterostructure of boron phosphide and SiC, and investigated the electronic, optical, and photocatalytic properties of the heterostructure with first-principle calculations.[111] The boron phosphide/SiC vdW heterostructure displayed a direct band gap with type II band arrangement, and showed an very high intensity of optical absorption of ~10$^5$ cm$^{-1}$ under the work conditions with a pH of 0-3. Figure 6c displayed the mechanism due to the band alignment for photocatalytic response of boron phosphide, SiC, and boron phosphide/SiC. The dashed line represented the oxidation-reduction potential of water decomposed at different pH values. The edges of VB and CB both reach the energy-favorable position and straddle on the oxidation-reduction potential of $H^+/H_2$ and $O_2/H_2O$, indicating the ability of boron phosphide/SiC heterostructure to split water under the working conditions of pH = 0-3. Based on first-principle density functional theory (DFT) calculations, Cui et al. studied the formation energy, thermal stability, band structure, electrostatic potential, charge transfer and optical behavior of MoSSe/SiC and WSSe/SiC vdW heterostructures.[55] Figure 6d presented a schematic diagram of photogenerated carrier migration in WSSe/SiC heterostructure, the photogenerated holes in WSSe transferred to the VB of SiC, and the photogenerated electrons in SiC migrated to the CB of WSSe. In addition, the characteristics of MoSSe and WSSe are very sensitive to strain, the tensile or compressive strain of the heterostructure can obviously modulate the charge transfer and regulate the band gap structure at the interface. The heterostructure can effectively utilize solar energy and be used as a high-efficient photocatalyst.

**3.3 Gas Sensors**

Gas sensors[116] are used in a wide range of applications, including intelligent systems, detection of hazardous and toxic gases, playing an important role in the safety of human life. The wide band gap, chemical inertness, and stability of SiC make it more suitable for high-temperature applications.[117] 2D materials are ideal and attractive candidates for gas sensing due to their high specific surface area and excellent properties.[118] Therefore, the gas sensors of 2D/SiC heterostructures are able to achieve high-performance at high temperature and various gas ambience.

Roy et al. demonstrated a gas sensor consisting of epitaxial graphene nanowalls (EGNWs)/SiC/silicon (Si).[119] The graphene-sheathed SiC nanowall was grown on a 3C-SiC seed layer, which provided nucleation sites and promoted anisotropic growth of SiC nanowalls. The metal/insulator/semiconductor (MIS) of EGNWs/SiC/p-doped Si based gas sensor achieved unprecedented sensitivity (82 µA/ppm/cm$^2$ of $H_2$) with a fast response shorter than 500 ms and the detection limit of 0.5 ppm. The gas sensing mechanism was attributed to the large number of edges and structural defects of vertically oriented nano-graphitic, which were acted as adsorption sites and facilitate rapid electron transfer. Besides, Lebedev et al. developed and optimized graphene/SiC gas sensors and biosensors by growing graphene films on semi-insulating 4H-SiC (Si face) substrates *via* thermal decomposition of SiC at ~1700 °C.[120] The gas sensor based on graphene/4H-SiC showed sensitivity to $NO_2$ concentrations at the level of 1 – 0.01 ppb, but lacked selectivity to determine the adsorbed molecules. The absorbed molecules on the surface of graphene can lead to changes in conductivity, but the opposite sign of resistance produced by some molecules may cause the total change to be close to zero.

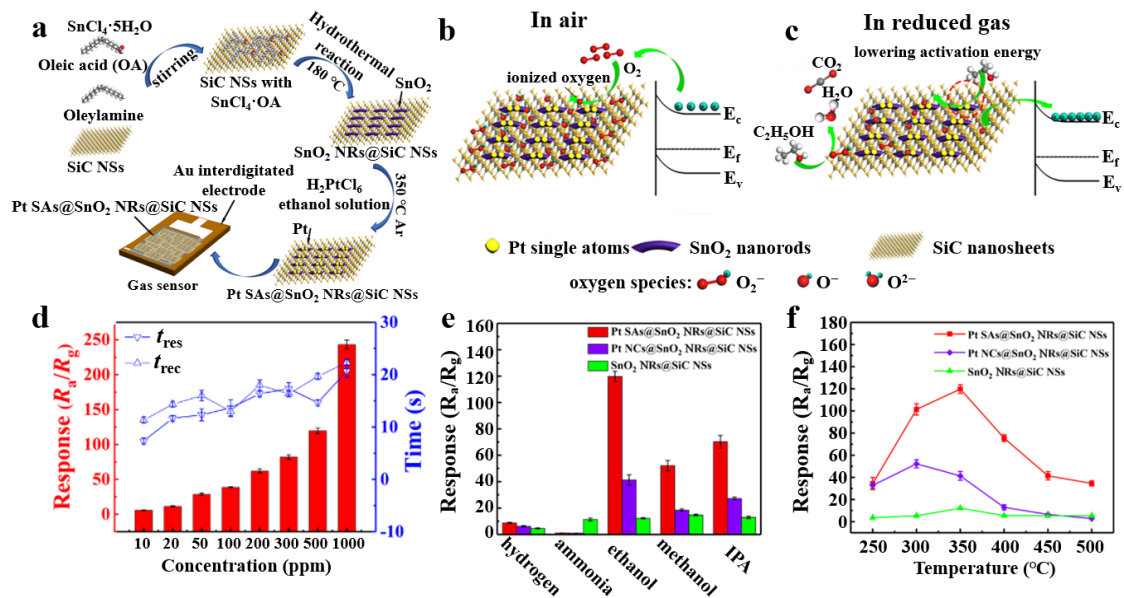

Figure 7 a. The preparation process of Pt single atoms (SAs)@SnO$_2$ NRs@SiC NSs. 1. SnCl$_4$·5H$_2$O, oleic acid and oleylamine were anchored on SiC NSs by stirring; 2. The mixture was converted into SnO$_2$ NRs@SiC NSs nanocomposites after hydrothermal reaction at 180 °C; 3. Pt single atoms were loaded onto SnO$_2$ NRs@SiC NSs nanocomposites by wet impregnation method using platinum precursor in argon atmosphere at 350 °C; 4. Pt SAs@SnO$_2$ NRs@SiC NSs gas sensor was fabricated by dipping the samples on a Au interdigitated electrode. Schematic illustration of gas sensing mechanism b. in air and c. in reduced gas of Pt SAs@SnO$_2$ NRs@SiC NSs. d. Response, response time t$_{res}$ and recovery time t$_{rec}$ toward concentrations of Pt SAs@SnO$_2$ NRs@SiC NSs under 500 ppm ethanol at 350 °C. Gas response of Pt SAs@SnO$_2$ NRs@SiC NSs, Pt NCs@SnO$_2$ NRs@SiC NSs, and SnO$_2$ NRs@SiC NSs, toward e. 500 ppm hydrogen, ammonia, ethanol, methanol, and IPA at 350 °C and f. 500 ppm ethanol vs. temperature. Reproduced with permission.[121]

Sun et al. constructed a new gas sensor based on the multi-heterostructure of Pt single atoms (SAs) @SnO$_2$ nanorods@SiC nanosheets (Pt SAs@SnO$_2$ NRs@SiC NSs) with the preparation process showing in Figure 7a.[121] The gas sensing mechanism was explained in Figure 7b and 7c, which was highly dependent on the surface reaction of the target gas and ionized oxygen. In the air, the gas sensing material absorbed oxygen molecules and converted to ionized oxygen species by electrons on CB, forming a depletion layer near the surface and leading to increased resistance. Under the action of the reducing gas, the gas molecules combined with oxygen species and transferred electrons to the sensing material, resulting in an increase in conductivity of the sensors. In Figure 7d, the response value increased with the increase of ethanol concentration, and the response time t$_{res}$ and recovery time t$_{rec}$ possessed a minimum value of ~14 and ~20 s, respectively. The gas responses of materials with different structures to various gas and temperature were shown in Figure 7e and 7f, indicating that Pt SAs@SnO$_2$ NRs@SiC NSs displayed highest gas sensing performance. The gas sensor showed excellent sensitivity with maximum response value $R_a/R_g$ ($R_g$ and $R_a$ represent the resistance of the sensor in target gas and air, respectively) of 119.75 ± 3.90, ppb-level detection, short response and recovery time (~14 and ~20 s), good selectivity and stability at high temperature of 500 °C. The large surface area, high electron mobility of 2D SiC nanosheets, and the tunable band gap of $SnO_2$ NRs further contributed to the fast gas-sensitive response. Each Pt atom in the multi-heterostructure acted as a catalyst to promote $O_2$ adsorption and dissociation and provide ionized oxygen, showing an unprecedented enhancement of gas-sensitive properties. Also, the $SnO_2$ NRs greatly facilitated the electron transfer process from the target gas to the sensor. The 2D nanostructure of SiC NSs provided a large surface area, abundant active sites, and sufficient electron transfer channels to enhance the sensing performance.

**3.4 Energy Storage Devices**

Advanced energy storage devices are being developed to meet the growing energy demand. The operating advantages of wide-band gap SiC in harsh environments such as high pressure and temperature are significant attributes for the development and application of energy storage devices.[122] The SiC-based heterostructure materials have potential for energy storage devices with high energy density and capabilities.

The heterostructure materials of graphene and SiC have received special attentions as potential electrode materials for electrochemical energy generation and storage. Emiliano et al. fabricated graphene-encapsulated SiC nanomaterials (graphene@SiC) the first time by simple adiabatic process, and used the heterostructures as supercapacitor materials and anodes in lithium-ion batteries (LIBs).[123] The reported graphene@SiC nanomaterials exhibited excellent supercapacitor behavior with a relatively high power density of 4800 W/kg and a specific capacitance of 394 F/g. The layered graphene@SiC device obtained a specific capacity of 150 mAh/g with a coulomb efficiency of ~99% and a current of 100 mA/g. Additionally, Liu et al fabricated highly crystalline graphene/3C-SiC nano-matrix on Si nanowires (G/SiC/SiNWs) for electric double-layer capacitors.[124] In Figure 8a, the G/SiC/SiNWs heterostructure exhibited a high area capacitance of 3.2 mF/cm$^2$ and a capacitance retention rate of 85% at a cyclic voltammetry (CV) scan rate of 50 mV/s and 100 mV/s, respectively. The inset in Figure 8a displayed the mechanism of electron transfer in the G/SiC/SiNWs nano-matrix heterostructure, the highly arranged nano-matrix heterostructure electrode provided more active sites for charge carrier storage and the fast ion transport pathways. In Figure 8b, the capacitance retention rate of G/SiC/SiNWs was increased by 115% after 10,000 test cycles with the CV scan rate of 100 mV/s, exhibiting good CV cycling stability and high specific capacitance. The enhancement could be attributed to the slow penetration of aqueous electrolytes into hydrophobic pores shown in the inset.

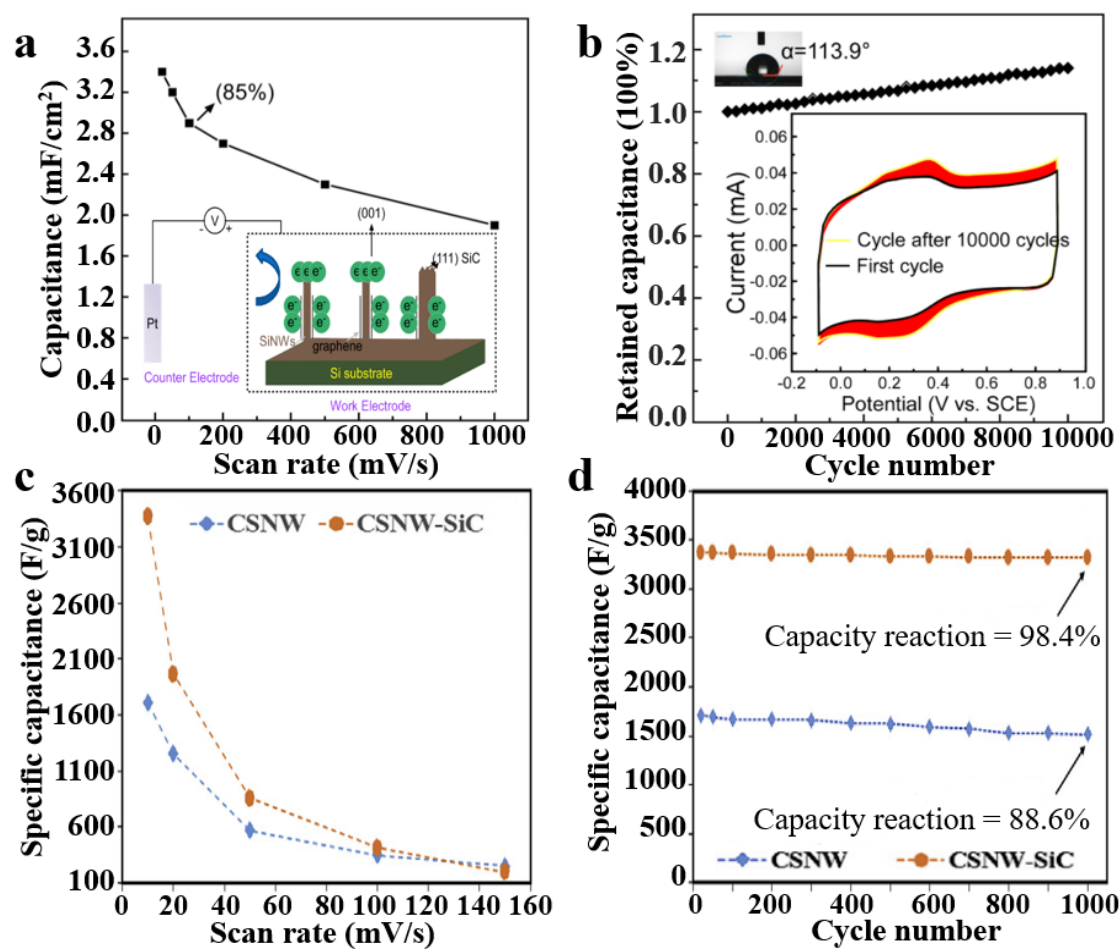

Figure 8 The capacitance of G/SiC/SiNWs nano-matrix a. at different CV scan rates and b. after the 10,000 cycles test. The inset in a is the schematic illustration of structure of the G/SiC on SiNWs nano-matrix (brown spikes) and the charge transferred from the SiNWs to SiC, then to outside circuit, and the insets in b show the cycling loops of CV and hydrophobic surface with water contact angle of 113.9°, respectively. Reproduced with permission.[124] The specific capacitances of CSNWs and CSNWs-SiC with c. different scan rates and d. cycle number. Reproduced with permission.[125]

More 2D/SiC heterostructure materials show potential application prospects in energy storage devices. For example, Zeraati et al. fabricated the SiC coated copper sulfide (CuS) nanowires (CSNWs) through chemical route, and enhanced the capacitive behavior and cycling stability of hybrid supercapacitors due to the unique electronic properties and chemical stability of SiC.[125] As shown in Figure 8c and 8d, the supercapacitance of CSNWs-SiC heterostructure enhanced from 1712 F/g to 3370 F/g, and the multiplicative performance increased from 88.6% to 98.4% after 1000 cycles. Xia et al. presented a self-assembled SiC/MXene heterostructure by ionization-bombardment-assisted deposition, providing a new method for the mass production of high-performance nanostructures and the fabrication of micro energy storage devices.[126] The SiC nanocrystals were obtained from the solid precursor powder under high pressure, then arranged into SiC nanowires and patterned into mesoporous SiC nanomesh to increase the specific surface area, and the obtained SiC nanomesh could serve as a framework to support the MXene films and provided a conductive network. The assembled SiC/MXene micro-supercapacitor had a specific capacity of 97.8 mF/cm$^2$ at current density of 1 A/cm$^2$, an energy density of 8.69 μWh/cm$^2$, and excellent multiplicative capability and stability with over 90% capacitance retention after 10,000 cycles at scan rate of 500 mV/s.

**3.5 Other Applications**

The applications of SiC-based 2D heterostructures go far beyond what has been mentioned above, their excellent performances also show great promises in other high frequency electronic devices and magnetic devices, and of course the fields that are unexplored and awaiting investigation.

For example, Li et al. presented a simple method for creating graphene/SiC tuned plasma cavities in the infrared (IR) range.[127] The absorption properties and field distribution of the resonance heterostructure were numerically investigated with a finite element method. At certain frequencies of SiC static scattering band, the structured SiC substrate acted as a perfect reflector to provide cavity effects *via* graphene plasma standing waves. Besides, Hu et al. carried out the research topic to explore the magnetic properties and stability of 2D magnetic materials *via* investigating the vdW heterostructure of CrI$_3$/SiC based on the DFT.[128] The ferromagnetic (FM) coupling of CrI$_3$ layer in the heterostructure can be improved with Curie temperature T$_c$ up to 62.3 K due to the Cr-I-Cr super-exchange interactions and proximity exchange effects. When T$_c$ increased to 93.14 K, the external electric field enhanced the FM coupling, the magnetic exchange energy increased steadily, and the interlayer gap decreased with T$_c$ reaching 128.8 K.

**4. Application of 2D Heterostructures Based on GaN**

The WBS material GaN[34], offers significant improvements in performance, efficiency, energy consumption and size compared to mainstream silicon power devices. The special crystal structure and wide energy gap of GaN, exhibit high power, high speed and high frequency characteristics in optoelectronic component applications. The high critical breakdown voltage and large saturation electron mobility make it potentially useful in power semiconductors and RF devices. However, low voltage operating environments, new packaging forms, heat dissipation requirements and excessive cost are the key issues that limit the further development of GaN-based devices. The combination of 2D materials and GaN can greatly improve the device performance by optimizing the energy band structure of the heterostructure. The advantages include the expansion of detection limit and faster charge transfer, etc., which contribute to more new application scenarios.

**4.1 Photodetectors**

GaN with the bandgap of 3.4 eV has high radiation hardness, visible blind detection capability, and ability to operate in both photovoltaic and photoconductive operational modes, making it a suitable semiconductor material for the development of efficient UV photodetector devices.[129] 2D materials usually possess a narrow bandgap, for example, $MoS_2$ and $WS_2$ with the bandgap of 1.8 and 2.0 eV, respectively.[130] Their combination is able to expand the detection limit from visible to ultraviolet light or from ultraviolet to visible, respectively. Furthermore, the GaN and typical 2D materials such as TMDs can form the type II band alignment, in which the conduction band minimum (CBM) and valence band maximum (VBM) of the two materials beside the interface are in staggered arrangement. The electrons and holes tend to be in lower energy state and are driven to the lower CBM or higher VBM and separate at the interface. Also, the built-in electric field at the interface of the heterostructure assists charge separation.[92] As shown in Figure 9, the minimum conduction band (CBM) and the maximum valence band (VBM) are located in GaN and $WSe_2$, respectively. The position of the Fermi level relative to the VBM suggests that the GaN epilayer and $WSe_2$ are almost intrinsic materials. The valence band offset $\Delta E_v$ and conduction band offsets $\Delta E_c$ of GaN/$WSe_2$ heterostructure are ~2.25 eV and ~0.8 eV, respectively. This kind of band offsets form type II energy band alignment at the interface, and these large offsets favor electron and hole separation, providing a way to integrate 3D III nitride materials and 2D TMDs materials for designing optoelectronic devices.[50]

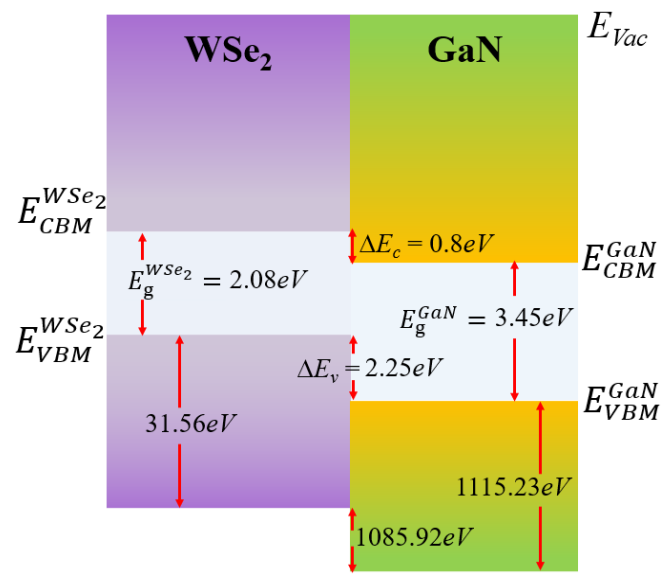

Figure 9 Schematic diagram of the type II energy band alignment at the interface of GaN/$WSe_2$.

For example, Li et al. reported the highly sensitive UV photodetectors based on monolayer graphene/nanoporous GaN heterostructure.[94] The device of nanoporous GaN with the highest porosity exhibited a fast and linear response to UV photons. The UV photodetector displayed good performance at low applied voltage −1.5 V, such as the ultra-high detectivity of ~1.0 × $10^{17}$ Jones and UV-visible rejection ratio of 4.8 × $10^7$. The responsivity and detectivity of nanostructure GaN-based UV photodetectors were much higher than graphene/bulk GaN due to their large specific surface area and nanoscale size.[131,132] Such high sensitivity was attributed to the significantly enhanced light collection and high mobility of nanoporous GaN, and the finite density of states of the monolayer graphene. In addition, Shen et al. fabricated the fast-response graphene/GaN UV photodetectors with optical response rise and decay times of 0.5 and 1.12 ms, respectively.[83] The methyl methacrylate as a support layer achieved ultra-clean graphene transfer, and greatly improved the performance of graphene/GaN photodetectors. Due to the effect of space charge region and the built-in electric field, the photogenerated holes and electrons flow to graphene and n-type GaN, respectively, resulting in a photocurrent. The transport of photogenerated carriers can further reduce the Schottky barrier, enabling a larger built-in electric field and faster hole-electron pair separation, thus generating in a large photocurrent.

In Figure 10a, Goel et al introduced the UV photodetector based on $MoS_2$/GaN heterostructure synthesized by wafer-scale sputtering method.[91] The great absorption of light by several layers of $MoS_2$ film led to high performance of $MoS_2$/GaN photodetector, showing high external spectral responsivity (~$10^3$ A/W) and detectivity (~$10^{11}$ Jones) with very fast response time (~5 ms), as displayed in Figure 10b and 10c. The observed responsivity and detectivity decreased with increasing light intensity due to the presence of trap states at the $MoS_2$/GaN interface and in the $MoS_2$ thin film. The photogenerated carriers are quenched by trap states at low light intensity, resulting in reduced recombination rate and extended lifetime of photoexcited carriers. However, an increase in light intensity leads to a decrease in the number of trap states, which leads to saturation of the photocurrent. In addition, Moun et al. investigated the surface potential of $MoS_2$/GaN heterostructure by Kelvin Probe Force Microscopy (KPFM) and confirmed type II energy band alignment.[92] The heterostructure device, as shown in Figure 10d and10e, demonstrated diode-like behavior attributed to the unique type II band alignment. In Figure 10f, the device appeared to be highly sensitive to 405 nm laser with the figure of merits of an ultra-high optical responsivity of $10^5$ A/W, a gain of $10^5$, and a detectivity of $10^{14}$ Jones. The decrease in responsivity and detectivity with increasing power density could be the presence of trap states at the interface of heterostructure. Besides, Zhang et al. reported $MoS_2$/GaN vdW heterostructure photodetectors (Figure 10g) based on large area homogeneous 2D-GaN flakes. The device showed a wide spectral responsive range from UV to visible region and enhanced photoresponse properties.[97] This is attributed to the good quality of 2D-GaN flakes and the fast separation of photogenerated electron-hole pairs driven by the built-in electric field in the depletion region of $MoS_2$/GaN p-n junction interface. As shown in Figure 10h, the photodetector revealed a high photoresponsivity of 328 A/W under 532 nm illumination. In Figure 10i, the device showed other meritorious parameters with EQE of 764%, specific detectivity of $2 \times 10^{11}$ Jones, fast response time of 400 ms under 532 nm laser.

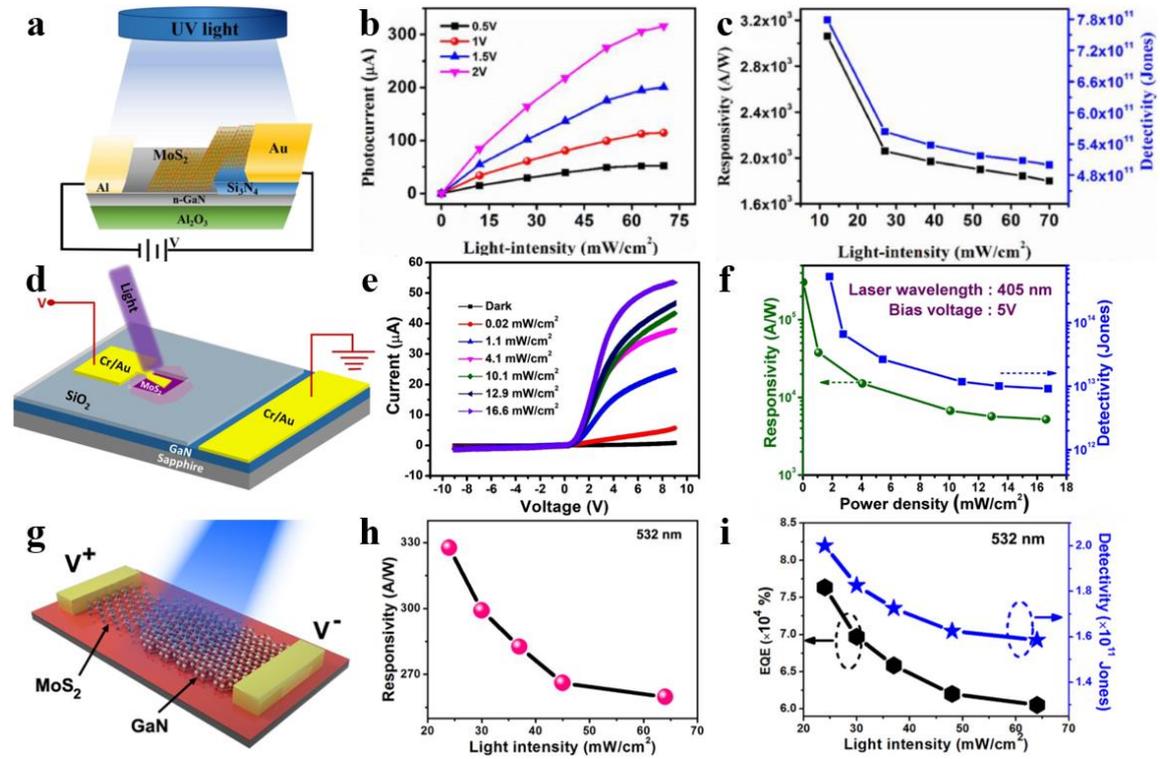

Figure 10 a. Schematic diagram of $MoS_2$/GaN heterostructure photodetector device; b. Photocurrent at different bias voltage and c. responsivity and detectivity at 1 V under different light intensity. Reproduced with permission.[91] d. Schematic illustration of $MoS_2$/GaN heterostructure device; e. Photoinduced I-V curves, f. Responsivity and detectivity under different laser intensity of $MoS_2$/GaN heterostructures. Reproduced with permission.[92] g. Schematic diagram of the $MoS_2$/GaN device; h. Responsivity, EQE and detectivity (i) of $MoS_2$/GaN at 532 nm illumination. Reproduced with permission.[97]

Furthermore, comparing the heterostructure photodetectors based on 2D/GaN and 2D/SiC, the described heterostructures of 2D/GaN exhibits better performance. The main parameters of photodetectors based on 2D/GaN and 2D/SiC heterostructures are listed in Table 3, and the responsivity of $MoS_2$/GaN displays much higher value than other heterostructures. Such excellent performances of $MoS_2$/GaN are attributed to the facts: (1) the interface effect such as built-in electric field and the type II band alignment of the heterostructures for MoS$_2$/WBS, which graphene/WBS does not possess. (2) GaN possesses a direct bandgap, unlike the indirect bandgap of SiC; (3) the band edge offsets are larger for MoS$_2$/GaN than MoS$_2$/SiC, resulting in larger potential for the charge transfer on the interface; (4) the electron mobility of GaN (1300 cm$^2$/V·s) is higher than SiC (900 cm$^2$/V·s of typical 4H-SiC).

Table 3 The performances of 2D/WBS heterostructure-based photodetectors at bias voltage $V_{bias}$ or gate voltage $V_g$.

| Heterostructures | Wavelength | Responsivity | Detectivity (Jones) | EQE (%) | Conditions | Ref. |
|---|---|---|---|---|---|---|
| graphene/6H-SiC | 325 nm | 2.18 A/W | — | 832 | $V_{bias} = -5$ V | [104] |
| graphene/4H-SiC | 325 nm | 254.1 A/W | 2.16×10$^{10}$ | 9.6×10$^4$ | $V_{DS} = -3$ V, $V_{GS} = 3$ V | [105] |
| graphene/4H-SiC | 365 nm | 189.7 mA/W | — | 66.6 | $V_{bias} = 10$ V | [106] |
| MoS$_2$/4H-SiC | 365 nm | 5.7 A/W | 3.07×10$^{10}$ | — | $V_{bias} = 20$ V | [107] |
| MoSe$_2$/WSe$_2$/4H-SiC | 532 nm | 7.17 A/W | 5.51×10$^{11}$ | 1.67×10$^3$ | $V_g = 10$ V | [108] |
| graphene/GaN | 360 nm | ~10 A/W | 1.0×10$^{17}$ | ~100% | $V_g = -1.5$ V | [94] |
| MoS$_2$/GaN | 365 nm | ~10$^3$ A/W | ~10$^{11}$ | — | $V_{bias} = 1$ V | [91] |
| MoS$_2$/GaN | 405 nm | 2×10$^5$ A/W | 6.0×10$^{14}$ | — | $V_{bias} = 5$ V | [92] |
| MoS$_2$/GaN | 532 nm | 328 A/W | 2×10$^{11}$ | 764 | $V_{bias} = 5$ V | [97] |

**4.2 Photocatalytic and Photoelectrochemical Water Splitting**

GaN-based nitrides show strong emission in the UV-visible region which is beneficial for spontaneous water splitting, such that GaN is also expected to be used as a photocatalyst or photoanode.[133,134] The heterostructure of GaN-based 2D materials is expected to show excellent photocatalytic or photoelectrochemical (PEC) properties.

For example, Chen et al. investigated the effect of nitride interfacial layer on the structure, electronic, and optical properties of MoS$_2$/GaN heterostructures by first-principle calculations and experimental analysis.[135] The MoS$_2$/GaN heterostructure with nitride interfacial layer formed by nitriding of Ga surface, can significantly improve the photocatalytic ability with boosted charge separation by band alignment and increased light absorption capability, showing excellent advantages in enhancing photocatalytic hydrogen production. Besides, Idrees et al. investigated the geometric, electronic, optical, and photocatalytic properties of MoSSe/graphene-like GaN (g-GaN) and WSSe/g-GaN heterostructures by first-principle calculations.[136] The heterostructures exhibited the type-II band alignment in all stacking modes, and proper valence and conduction band edge positions. The standard redox potential provided sufficient forces to drive the photogenerated electrons and holes to dissociate water into $H^+/H_2$ and $O_2/H_2O$ at pH=0. Sun et al designed a novel 2D material vdW heterostructure composed of g-GaN and boron selenide (BSe) as an efficient photocatalyst for water decomposition.[137] Figure 11a-c showed the band structure of g-GaN/BSe vdW heterostructure, which had a stable configuration with the binding energy of −54.36 meV/Å$^2$ and showed thermal stability at room temperature. The type II band alignment of g-GaN/BSe continuously promoted the separation of photogenerated electron-hole pairs and provided a suitable band edge for the redox potential of water decomposition at pH of 0 and 7. The excellent daylight absorption can be further enhanced by biaxial strain.

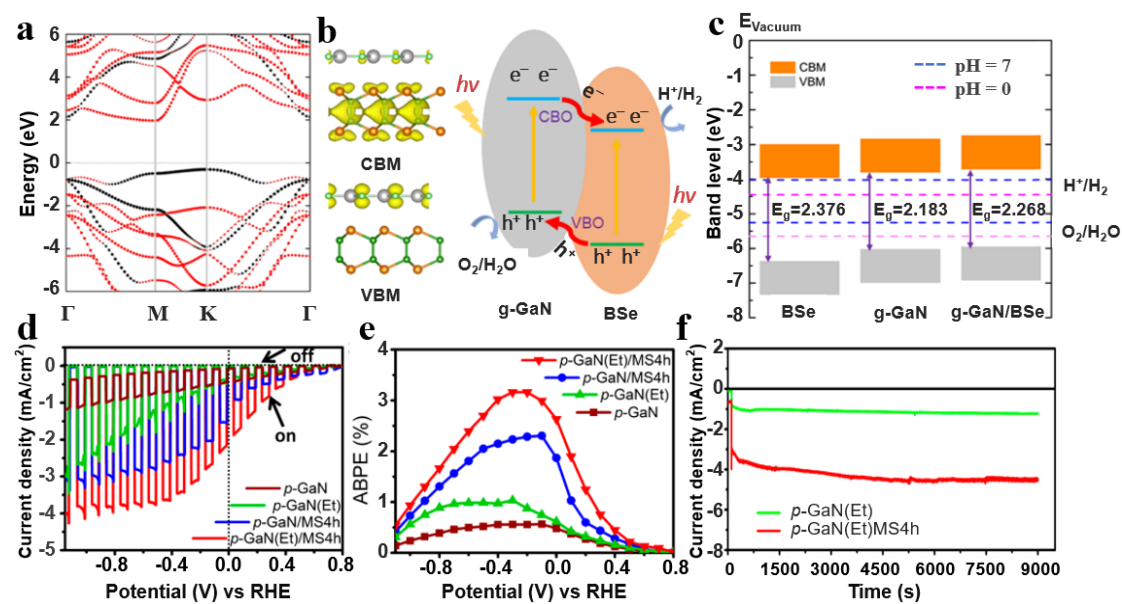

Figure 11 a. Predicted energy band structure of g-GaN/BSe; b. (left) Band-resolved charge densities of VBM and CBM of g-GaN/BSe heterostructure, the gray, light green, green and orange spheres indicating Ga, N, B and Se atoms, respectively; and (right) schematic diagram of photogenerated charge migration at the g-GaN/BSe interface; c. The band edge positions of monolayer g-GaN, BSe and the g-GaN/BSe heterostructure. Reproduced with permission.[137] d. Linear sweep voltametric curve under chopped light and e. applied bias photon-to-current conversion efficiency (ABPE) for pristine p-GaN, p-GaN(Et), p-GaN/MS4h, and p-GaN(Et)/MS4h; f. Stability *vs.* time under continuous illumination of p-GaN(Et), p-GaN(Et)/MS4h photocathode. Reproduced with permission.[96]

Ryu et al. reported a novel heterostructure of $MoS_2$/GaN as a photoanode for solar-driven PEC water splitting.[138] The $MoS_2$/GaN photoanode achieved efficient light collection with the photocurrent density of 5.2 mA/cm$^2$ at 0 V, which was 2.6 times of the value of original GaN. The current conversion efficiency of 0.91% of $MoS_2$/GaN was higher than that of GaN (0.32%). Furthermore, Kumar et al. demonstrated the further modification of chemical etched p-GaN (p-GaN(Et)) nanotextured hexagonal micropores by 2D-$MoS_2$, resulting in an interconnected nanowall networks that can be used for PEC hydrolyzed photocathodes (PCs).[96] Figure 11d-f presented the PEC performance of 2D-$MoS_2$/p-GaN(Et): after 4 hours decoration of $MoS_2$ (MS4h), the photocurrent increased from −0.43 mA/$cm^2$ (p-GaN) to −1.52 mA/$cm^2$ (p-GaN/MS4h). The optimized photocathode displayed a maximum applied bias photon-to-current conversion efficiency (ABPE) of ~3.18% at −0.3 V *vs.* reversible hydrogen electrode (RHE), 3 times higher than that of the p-GaN(Et) electrode (~1.03%). The calculated hydrogen evolution rate increased from 43.6 μmol/h of p-GaN(Et) photocathode to 89.56 μmol/h of p-GaN(Et)/MS4h. The p-GaN(Et)/MS4h demonstrated an exceptional stability under continuous operation at 1 solar illumination (100 mW/$cm^2$). The heterostructures of GaN and TMDs materials have great prospects toward efficient and stable photocathodes for the PEC water splitting.

### 4.3 Light Emitting Diodes

GaN-based light emitting diodes (LEDs) have been extensively studied, but most of the energy are dissipated in form of heat despite the increasing efficiency of devices, especially the blue light LEDs based on InGaN/GaN.[139,140] The GaN LED devices embedded with graphene oxide are capable of emitting bright light at low temperature and thermal resistance.[141] 2D materials such as graphene oxide alleviates the self-heating problems, making the device performance better than conventional ones. Therefore, the 2D/GaN heterostructures are expected to show good performance in LED devices. For example, Lin et al. introduced the high-performance surface plasmon-enhanced graphene/p-GaN LEDs by inserting Ag nanoparticles (Ag NPs) into the graphene/p-GaN interface, which showed broadband emission from 550 nm to 650 nm under forward bias and sharp emission at ~400 nm under a biased voltage.[142] Besides, Su et al. fabricated the inorganic p-GaN/n-$SnO_2$ hetero-diodes that can be used to develop solar cells, dual-color LEDs, and self-powered devices.[143] The p-GaN/n-$SnO_2$ heterodiode displayed an extremely high responsivity of 185 mA/W, a high EQE of ~74%, ultrahigh response speed with the rise time $t_r$ of 340 ns and decay time $t_d$ of 61 μs, and a high UV-visible rejection ratio of ~$2.1 \times 10^3$. Also, a linear dynamic range (LDR) of 87 dB, an on/off ratio of $2.1 \times 10^4$, a specific detectivity of $2.81 \times 10^{13}$ Jones at small bias voltage of −0.01 V were achieved.

### 4.4 Sensors

GaN nanostructure-based gas sensors such as nanowires (NW), nanoparticles (NP), and nanonetworks (NN) have been reported for the detection of a variety of gases such as hydrogen, alcohols, methane, benzene, and their derivatives.[144] Although most of the reported GaN nanostructure-based gas sensors show promising performance, many challenges remain in terms of device sensitivity, selectivity, response/recovery speed, and reliability. In order to improve these figures of merits, people start to incorporated 2D materials with ultrahigh specific area and large chemical tunability together with GaN. For example, Kumar et al. fabricated the $MoS_2$/GaN heterostructure-based gas sensors by magnetron sputtering and sulfidation processes.[90] The sensing mechanism of the gas sensor was shown in Figure 12a, the electrons transferred from GaN to $MoS_2$, resulting in the energy band bent downward near the $MoS_2$ surface and bent upward near the GaN surface. The height of the interface barrier increased due to the adsorption of hydrogen at the interface, and the raised barrier structure was illustrated by the dotted red line, leading to a decrease in current across the junction and showing a very high sensitivity. The significantly improved sensitivity of gas sensor is shown in Figure 12b the sensitivity of $MoS_2$/GaN was several orders of magnitude higher than conventional only $MoS_2$- or GaN-based sensors, which increased from 21% to 157% for 1% hydrogen with temperature increased from 25 °C to 150 °C. According to the change in barrier height under hydrogen in Figure 12c, the change in barrier height increased with the increase of $H_2$ concentration and reached its maximum value at 150°C. Moreover, Kim et al. successfully fabricated the nitrogen oxide ($NO_x$) gas sensors of n-$MoS_2$/p-GaN (Mg doped) heterostructure with a response of 48.42 % to 50 ppm $NO_2$, which was 20 times of the value of bare p-GaN.[84] The energy band diagram in Figure 12d was used to discuss the gas sensing mechanism, the adsorbed $NO_x$ extracts electrons from $MoS_2$ and O ions and converts into $NO_x^-$, resulting in a decrease in the electron density at the interface of the heterostructure. The Fermi level $E_f$ of $MoS_2$ moves downward, and the barrier height of the heterostructure interface increases. Figure 12e shows that the sensor response enhanced to 98.77% for 50 ppm $NO_2$ in UV irradiation. And Figure 12f shows a response increased from 47.48% to 148.81% with the temperature increased from 27 to 150 °C due to the thermal activation of charge transfer.

Furthermore, Lin et al fabricated a $MoS_2$/Au/GaN based PEC aptamer sensor for the detection of cancer biomarker alpha-fetoprotein (AFP).[145] An aptamer is a screened oligonucleotide fragment that can bind to the target with high selectivity. The aptamers of AFP were modified on the Au/GaN photoelectrode surface *via* Au-S bonds and bonded to the target protein with high selectivity, thus preventing the charge transfer and reducing the suppression of photocurrents.[146] The Au film promoted charge transfer from GaN to $MoS_2$ and lead to a greater complexation of electrons and holes. This method can be used to detect the significantly suppressed photocurrents. The difference in photocurrent with and without AFP was linearly related to AFP concentration, and in the range of 1.0 - 150 ng/mL with a detection limit of 0.3 ng/mL. The combination of $MoS_2$/Au/GaN and aptamer can be applied to develop novel photoelectrochemical aptamer sensor for the model cancer biomarkers AFP.

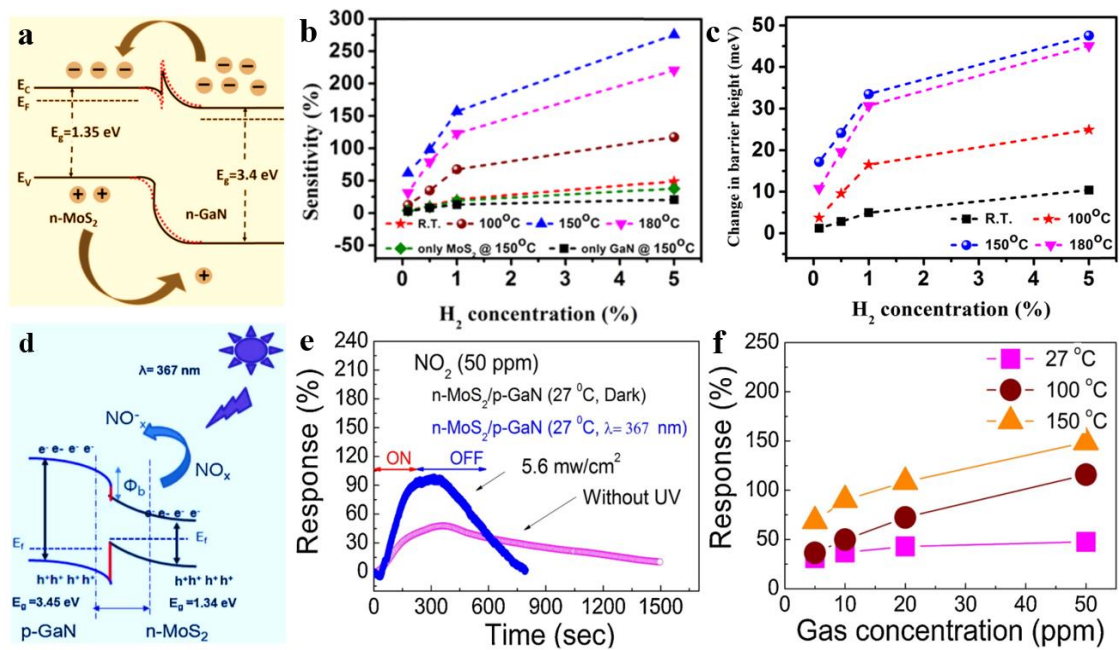

Figure 12 a. Schematic illustration of the energy band of the $MoS_2$/GaN heterostructure. b. Sensitivity vs. $H_2$ concentration and c change in barrier height at different temperatures of $MoS_2$/GaN heterostructure sensor. Reproduced with permission.[90] d. Schematic illustration of the gas sensing mechanism of n-$MoS_2$/p-GaN heterostructure in $NO_x$ ambience. e. The response vs. time with and without UV illuminations at 27 °C and f. responses vs. gas concentration at different temperatures of n-$MoS_2$/p-GaN sensor in $NO_2$ ambience. Reproduced with permission.[84]

## 4.5 Other Applications

The unique electronic interaction and good geometric matching structure between GaN and 2D materials provide ideal electron migration channels for high charge carrier extraction efficiency, resulting in excellent performance and showing brilliant promise for multiple applications.

For example, Huang et al. described a novel memory device based on wrinkled $MoS_2$-GaN nanowires (NWs), showing good stability and retention characteristics of 798 cycles and $3.4 \times 10^3$ s with a low switching voltage.[147] Defects may be formed in the wrinkled structure, causing O ions to migrate along the wrinkled surface of NWs, and resulting in the memristive behavior of the device. Moreover, Chen et al. demonstrated the simultaneous spontaneous four-wave mixing (SpFWM) in the visible spectral range, which was driven by photoluminescence from the single layer $MoS_2$ clad on GaN micro-disk cavity.[148] The monolayer $MoS_2$ thoroughly absorbed the resonant external pump due to the efficient round-trip evanescent coupling, and achieved a powerful pump for SpFWM generation through the PL emission. The output spectral distribution was deeply depended on probe position, drive power and temperature. The PL-induced SpFWM was performed in the visible light range for the first time, due to the effective feedback of the microdisk cavity and the strong interaction between the lattice-matched $MoS_2$ and GaN. Furthermore, Desai et al. fabricated the $MoS_2$-GaN heterostructure diodes, showing excellent diode characteristics with significantly low reverse saturation current density of $1.37 \times 10^{-6}$ $A/cm^2$ at $-2$ V and a rectification ratio of $5 \times 10^7$.[49] Photovoltaic effects were also observed under light illumination at wavelengths of 300 – 1100 nm, important range for photosensitive device applications.

## 5. Conclusions and Perspectives

The heterostructures of 2D/WBS materials provide new platforms for theoretical studies and experiments in fundamental physics. They hold excellent properties that single-phase 2D or WBS materials do not possess, leading to new physical phenomena or significant improvement of the performance of devices. Here, we briefly introduce the fabrication methods of 2D/WBS heterostructure materials, including the direct preparation methods such as CVD and PLD, as well as the most widely used multi-step construction methods combining chemical or physical approaches of exfoliation and transfer or etching. Furthermore, we mainly focus on the applications of 2D/WBS heterostructure materials, covering a variety of topics, including photodetectors, photocatalysis, photoelectrochemical, photodiodes, novel sensors, energy storage devices, and memristor devices. The interface effects of the heterostructures, such as band alignment, built-in electric field, defects, doping, or super-exchange interaction at the interface, have a crucial influence on the performance of the device. The various 2D/SiC and 2D/GaN heterostructures presented in this review demonstrate the diversity of applications of 2D/WBS heterostructures, having the potential to develop practical high-performance devices. The results achieved are briefly summarized as follows.

For 2D/SiC heterostructures, current researches on 2D/SiC primarily focus on graphene/SiC heterostructures, followed by heterostructures of SiC with TMDs and MXene, etc. The photodetectors of graphene/SiC heterostructures exhibited favorable performances, such as high photoresponsivity, EQE and detectivity. A variety of 2D/SiC heterostructures have been used as high-efficiency photocatalysts. For example, $SnO_2$/SiC heterostructure device demonstrated a high hydrogen evolution rate and large current density at low voltage. Besides, the gas sensor of EGNWs/SiC/Si achieved unprecedented sensitivity with fast response, and even the supercapacitors of CuS/SiC heterostructure for energy storage displayed a high capacitance.

For 2D/GaN heterostructures, they exhibited obvious advantages for the development of optoelectronic devices, especially the $MoS_2$/GaN heterostructures due to the excellent properties of GaN and the interface effects. The photodetectors of different $MoS_2$/GaN heterostructures showed high responsiveness and detection in the UV-visible range. The heterostructures of TMDs/GaN also displayed excellent promise in photocatalysis and photoelectrochemical hydrolysis. In addition, multiple types of devices such as the dual-color LEDs of n-$SnO_2$/p-GaN heterostructure, high sensitivity gas sensors and novel PEC aptamer sensor of $MoS_2$/GaN heterostructures, revealed the great application potential of 2D/GaN heterostructures.

Although theoretical predictions or experiments have been reported and confirmed on 2D/WBS heterostructures, the researches on them are still in the preliminary stage. Both fundamental researches and practical devices on 2D/WBS heterostructures still face many challenges but also opportunities. These issues need to be addressed for further applications, such as the quality of heterostructures, interface modulation mechanisms, and device paradigms of the 2D/WBS heterostructures.

(i) The functionalities largely rely on the quality of heterostructures, including the quality of materials and the interface. The exfoliation and transfer will induce residues. CVD has problems of controllability and stability. PLD yields polycrystalline 2D materials. Moreover, the combination of multiple methods is complicated and difficult to integrate high-quality 2D/WBS heterostructure devices. How to solve the scalable synthesis of 2D materials and achieve high-quality heterostructures, especially in a large scale? It awaits more investigations.

(ii) It is crucial to explore new physics, realize multi-platform functions and study the mechanisms of modulation on the optical, electrical or magnetic properties of 2D/WBS heterostructures in more depth. The unique properties of 2D materials, such as strong spin-orbit coupling, spin-valley locking, high layer number and interfacial tunability, topological physical properties are all not exploited for 2D/WBS heterostructures yet. Besides, the high breakdown electric field, thermal conductivity, and electron velocity of WBS materials could be added to the study of 2D-WBS heterostructures in high-power and microwave devices. Novel physics and applications could be achieved by taking into account of these.

(iii) Designing and implementing high-performance and high-uniformity device patterns is the most challenging in device applications. Currently, the typical integration method of 2D and WBS materials is by transfer. The interface quality is influenced by the wrinkles, disorder and the organic residue induced by the transfer procedure. The organic residue will also produce some doping effect. This integration method is time consuming and in low efficiency and scalability. Questions such as how to break through the key scientific issues of heterostructure integration, and achieve effective and scalable integration and interconnection of 2D/WBS devices are also urgent for answers.

The integration of 2D/WBS heterostructure materials will combine the unique properties of both, promises to provide not only flexibility in constructing hybrid structure with a variety of material choices but also novel functions that conventional devices have inadequate capability to achieve. Moreover, the study of interface doping or defects on specific 2D channels is critical for miniaturizing devices in integrated circuits.[149] From the perspective of miniaturization in integrated circuits, when the channel size becomes comparable to the size of the potential spatial fluctuations, the impact on device operation is very strong. Despite the growing understanding of interface properties in 2D systems, the detailed study and analysis of the interfacial properties of 2D heterostructures is still quite limited. It is believed that this exciting field would open a new avenue for realizing the next generation of modern electronics and optoelectronics.

**Acknowledgements**

The authors acknowledge support from the National Natural Science Foundation of China (Grant Nos. 51902061, 62090031, and 91964107).

**Conflict of Interest**

The authors declare no conflict of interest.